\documentclass[preprint,aps,prd,showpacs,showkeys,amsmath,amssymb,eqsecnum]{revtex4}

\usepackage{dcolumn}
\usepackage{bm}

\begin{document}

\preprint{Nice INLN 07/18 July 2007, Brown-HET-1498}

\title{Bloch-Nordsieck Estimates of High-Temperature QED}


\author{H. M. Fried$^\star$, T. Grandou$^\dagger$ and Y.-M. Sheu$^\star$}
\affiliation{$^\star$ Brown University, Physics Department, Box
1843,Providence, RI 02912, USA \\ $^\dagger$ Institut Non
Lin$\acute{e}$aire de Nice UMR CNRS 6618; 1361, Route des Lucioles,
06560 Valbonne, France.}


\date{\today}

\begin{abstract}
In anticipation of a subsequent application to QCD, we consider the
case of QED at high temperature. We introduce a Fradkin
representation into the exact, Schwingerian, functional expression
of a fermion propagator, as well as a new and relevant version of
the Bloch-Nordsieck (BN) model, which extracts the soft
contributions of every perturbative graph, in contradistinction to
the assumed separation of energy scales of previous
semi-perturbative treatments. Our results are applicable to the
absorption of a fast particle which enters a heat bath, as well as
to the propagation of a symmetric pulse within the thermal medium
due to the appearance of an instantaneous, shock-wave-like source
acting in the medium. An exponentially-decreasing time dependence of
the incident particle's initial momentum combines with a stronger
decrease in the particle's energy, estimated by a sum over all
Matsubara frequencies, to model an initial "fireball", which
subsequently decays in a Gaussian fashion. When extended to QCD,
qualitative applications could be made to RHIC scattering, in which
a fireball appears, expands and is damped away.
\end{abstract}

\pacs{11.10.Wx, 11.15.Tk, 12.20.Ds}
\keywords{Finite temperature, Bloch-Nordsieck, soft photon, Fermion damping, QED,
Matsubara formalism, Matsubara frequency, functional method, plasma.}

\maketitle

\section{\label{sec1}Introduction}

In a previous article \cite{CFG2005}, a hot quantum field toy model
was used in order to test and appreciate the calculational
efficiency of the functional methods long developed and used by one
of us in a large variety of situations (for example, Refs. \cite{HMF1972,HMF1990,HMF2002}). The result
for the two-point function came out both non-trivial and remarkably
simple, opening on some interesting physical interpretations. The
scalar model, however, had little to connect it to the physical
theories of QED and QCD. In the case under consideration of QED, and
at the same level of approximation, the same two-point function
exhibits a far richer structure of entwined contributions and
associated mechanisms, which are the matter of the present article.

The mechanisms for depletion of a high-energy particle's energy $E$
and momentum $p$ when incident upon a medium at equilibrium
temperature $T$, where $p \gg T$, suggested in the following
sections are intended to be a small improvement to the seminal work
of Weldon \cite{Weldon1991a}, Takashiba \cite{Takashiba1996a}, and Blaizot and Iancu \cite{BI1996a,BI1997a,BI1997b} of a decade ago.
Our techniques and points of view are somewhat different from
theirs, but it is essential to begin by acknowledging our debt to
these authors, who first introduced and implemented the idea of a Bloch-Nordsieck
(BN) approximation in order to estimate the Physics of such energy
and momentum depletion.

It may be useful to note the interpretations that we bring to this
subject, and we here enumerate the special aspects of our approach
which the interested reader will encounter below.

1. At the very beginning, we separate and discard (the infinities
of) those aspects of free-particle mass and wave-function
re-normalization from the specific effects of the medium on the
particle. This is simple to perform in a functional approach, but
rather complicated in the conventional, momentum-space expansion of
the proper self-energy part of the inverse fermion propagator (of
the particle which is entering the medium). Specific effects of the
effective mass change in this model due to the motion of the
particle in the medium may be found in the Ph.D. thesis of one of
us \cite{Sheu2008a}.

2. We introduce a modified BN approximation appropriate to the case
when the particle's momentum is decreasing as it moves into the
medium. Energy losses are calculated by the usual, Matsubara
replacement of $E \rightarrow \omega_{n}$, and thermally averaged
according to the Martin-Schwinger/Matsubara formalism we use, as
described in detail in a previous publication \cite{CFG2005}. But
momentum loss is described by a separate, "Doppler" mechanism, which
replaces the constant BN momentum $p$ by $p(t) = p(0) \exp{(-\Gamma
t)}$, where $t$ is the duration of time the particle has been in the
medium, and $\Gamma$ is specified by a simple, semi-classical
argument. We feel this choice of semi-classical BN momentum is more
physical than the conventional procedure, appropriate to high-energy
scattering, of retaining the constant value which the particle has
upon entering the medium; but we place no particular emphasis on
this Doppler mechanism for calculating $p(t)$. Other models may well
be better for the description of this decreasing BN parameter, but
this one is simple, and physically reasonable, and has the
interesting consequence of modeling the appearance of a "fireball"
at the initial stages of the particle's thermal history.

3. We view the medium as an effective mechanism for the loss of a
particle's energy and momentum, without requiring the particle to
remain continuously on its mass shell (this is good Quantum
Mechanics, because the experiments we are describing do not measure
this property!). Only after thermalization, when $p(t)$ has
decreased to the order of $T$, and its derived exponential decay
law is no longer relevant, only when the particle joins its many
identical twins in the equilibrium distribution at temperature $T$,
can the particle be supposed to be on its mass shell.

 4. We rigorously maintain the nature of our BN approximation, with
all real or virtual $k_{\mu}$ coupled to the incident particle
required to satisfy $|\vec{p}| \gg |\vec{k}|$. As a result, all integrals are
finite, and easily approximated.  For reasons stated in Section \ref{sec3} (after Eqs. (\ref{Eq3-11}) and (\ref{Eq3-16})), we do not employ the conventional HTL analysis to describe pair-production generated by virtual
photons emitted in the medium by the incident particle; rather, we
estimate such pair-production using a straight-forward functional
representation, and find it multiplying the ordinary Bremsstrahlung (a contribution to the decay exponent of $g^{2} \, (\vec{p}^{\, 2})^{2}$) by a factor of $g^{2} \ln{(\vec{p}^{\, 2}/m^{2})}$.  Were the coupling large, rather than that of QED, this term could be suppressed by the unitary denominator factor as described after Eq. (\ref{Eq3-16}).

5. We are able to provide an explicit expression for the
time-dependence of the thermalization process, as the particle's
$|\textrm{thermal average propagator}|^{2}$ initially increases
-corresponding to the "fireball" -and then decreases rapidly, as
given by a specific, Gaussian decay. Were we to restrict the final
Matsubara sum to $n = 0$ only, that fall-off would be exponential;
but we are able to sum over all $n$, and the result is a stronger,
Gaussian approach to thermalization.

6. Our model calculation is able to distinguish longitudinal and transverse
components of the "fireball".  We do not actually compute
distributions which resemble a true fireball; rather, we use the
word to represent a short-lived enhancement of probability as a
function of time in the medium, corresponding to the incident
particle's ability to generate a longitudinal burst of secondary
particles and photons. By "transverse fireball" is meant a
short-lived enhancement of probability as a function of the incident
particle's time in the medium, which can serve to generate a
symmetric pulse of secondaries in an arbitrary direction.

 The paper is organized as follows. Section \ref{sec2} describes the
essential features of our BN derivation for the fermionic two-point
function, in quenched approximation.  The result turns out to be
remarkably simple. For the sake of completeness, the theoretical
steps which come before that treatment are deferred to Appendix
\ref{appA}. Then, a "Doppler" model for the fast particle momentum
damping is used to conclude Section \ref{sec2} and the Doppler
model itself described in Appendix \ref{appB}. In Section
\ref{sec3}, the approximation of quenching is removed so as to take
fermionic loop leading effects into account.  The transverse fireball will be discussed in Section \ref{sec4}.  A summary and a discussion of our results are presented in Section \ref{sec5}.

\section{\label{sec2}Quenching within the Bloch-Nordsieck approximation
scheme}

The main steps of the approach are as follows, as succinctly  as
possible.

Inherent to the BN approximation scheme, ordered exponentials which
appear in the rigorous Fradkin representation of the fermion
propagator \cite{HMF1990,HMF2002},
\begin{eqnarray}\label{Eq2-1}
& & \left( e^{g \int_{0}^{s}{ds' \, \mathbf{\sigma} \cdot \mathbf{F}(y-u(s'))}}
\right)_{+}, \\ \nonumber & & \quad \mathbf{\sigma}_{\mu \nu} = \frac{1}{4}\left[\gamma_{\mu},
\gamma_{\nu} \right], \quad \mathbf{F}_{\mu \nu}=\partial_{\mu} A_{\nu} -
\partial_{\nu} A_{\mu}
\end{eqnarray}

\noindent are suppressed, because, as in all eikonal/BN models, they
generate terms proportional to soft photon momenta, which can be
neglected in comparison to the particle's momentum. Because of our
suppression of conventional mass renormalization, and the understood
appearence of spinorial wave functions on either side of the final
propagator, the fermionic propagator at zero temperature will have
its $(m - i \gamma \cdot p)$ factor replaced by $2m$, and will read
\begin{eqnarray}\label{Eq2-2}
& & \langle x_{0}, \vec{p}|\mathbf{S}_{c}|y_{0}, \vec{y} \rangle = \\ \nonumber & & i \, (2m)
\, e^{-i \vec{p} \cdot \vec{y}} \, \int_{0}^{\infty}{ds \, \int{dp_{0} \, e^{-is(\omega^{2} - p_{0}^{2})} \, e^{- i
p_{0} (x_{0}-y_{0})} }},
\end{eqnarray}

\noindent where $\omega^{2} = \vec{p}^{2} + m^{2}$.  With the same approximations, the free-fermion thermal
propagator is given by
\begin{eqnarray}\label{Eq2-3}
\nonumber \tilde{\mathbf{S}}_{th}(\omega, z_{0}) &=& (2m) \, \frac{i}{\tau}
\sum_{n=-\infty}^{+\infty}{ \int_{0}^{\infty}{ds \,
e^{-is(\omega^{2} - \omega_{n}^{2})} \, e^{-i \omega_{n} z_{0}} }},
\\ \omega_{n} &=& \frac{(2n+1) \pi}{\tau}, \quad z_{0} = x_{0} -
y_{0},
\end{eqnarray}

\noindent and, by direct evaluation, is equal to
\begin{eqnarray}\label{Eq2-4}
& & \tilde{\mathbf{S}}_{th}(\omega, z_{0}) = \\ \nonumber & & \quad \quad (2m) \, \frac{i}{2 \omega}
\left\{ [1 - \tilde{n}(\omega)] e^{- i \omega z_{0}} -
\tilde{n}(\omega) e^{+i \omega z_{0}} \right\},
\end{eqnarray}

\noindent for $z_{0} > 0$, where $\tilde{n}(\omega)$ is the Fermi-Dirac distribution function \cite{LeBellac2000}.  For the thermal propagator in the presence of a background $A_{\mu}$-field, the corresponding BN approximation gives
\begin{eqnarray}\label{Eq2-5}
& & \langle x_{0}, \vec{p}|\mathbf{G}_{th}^{BN}[A]|y_{0}, \vec{y} \rangle = \frac{i}{\tau} \sum_{n=-\infty}^{+\infty}{ \int_{0}^{\infty}{ds }} \\ \nonumber & &  \quad \quad \quad \times {e^{-is(\omega^{2} - \omega_{n}^{2})} \, e^{-i \omega_{n} z_{0}} \,
e^{-ig \int_{0}^{s}{ds' \, p \cdot A(y - 2 s' p)} }},
\end{eqnarray}

\noindent where $p_{\mu} =(\omega_{n}, \vec{p})$. The thermal
2-point fermion function, in quenched BN approximation, is then given
by
\begin{widetext}
\begin{equation}\label{Eq2-6}
{\langle \vec{p}, n| {\mathbf{S}'}_{th}^{BN} |\vec{y}, y_{0}
\rangle} = \left. e^{-\frac{i}{2} \, \int{\frac{\delta}{\delta A_{\mu}} \,
\mathbf{D}_{th}^{\mu \nu} \, \frac{\delta}{\delta A_{\nu}}}} \cdot
{\langle \vec{p}, n| \mathbf{G}_{th}^{BN}[A] |\vec{y}, y_{0} \rangle}
\right|_{A \rightarrow 0} \cdot \mathbf{Z}_{0}[i\tau],
\end{equation}
\end{widetext}

\noindent where $\mathbf{Z}_{0}[i\tau]$ is the free partition function, and in the real-time imaginary-temperature formalism
being used for the linkage operator with Matsubara sum, one has $\mathbf{D}_{th}^{\mu
\nu} = \mathbf{D}_{c}^{\mu \nu}  + \delta \mathbf{D}_{th}^{\mu \nu}
$, with $\mathbf{D}_{c}^{\mu \nu}$ the causal free-photon
propagator, and $\delta \mathbf{D}_{th}^{\mu \nu}$ the proper
thermal part of $\mathbf{D}_{th}^{\mu \nu}$,
\begin{widetext}
\begin{eqnarray}\label{Eq2-7}
\delta \mathbf{D}_{\mu \nu}^{th}(u-v) &=&  \frac{i}{(2 \pi)^{3}}
\int{d^{4}k \, \delta(\vec{k}^{\, 2} - k_{0}^{2}) \,
\frac{e^{i k \cdot (u-v)}}{e^{\beta |k_{0}|} - 1} \, \hat{\mathbf{D}}_{\mu \nu}} \\
\nonumber &=&  \frac{i}{(2 \pi)^{3}} \int{\frac{d^{3}\vec{k}}{2k} \,
\frac{e^{i \vec{k} \cdot (\vec{u} - \vec{v})}}{e^{\beta k} - 1} \,
\, \left( e^{-i k_{0} (u_{0} - v_{0})} + e^{+i k_{0} (u_{0} -
v_{0})} \right) \, \hat{\mathbf{D}}_{\mu \nu}}
\end{eqnarray}
\end{widetext}

\noindent with $k = |\vec{k}|$. In the Coulomb gauge to be used, one
has $A_{0} = 0, \nabla \cdot \vec{A} = 0$, and
\begin{equation}\label{Eq2-8}
\hat{\mathbf{D}}_{\mu \nu} = \delta_{\mu i} \, \left(\delta_{i j} -
\frac{k_{i} k_{j}}{\vec{k}^{2}} \right) \, \delta_{j \nu}.
\end{equation}

In this first, quenched approximation, all complications related to
conventional, $T=0$, free-particle mass and wave-function
re-normalization are removed by suppressing the $\{ -\frac{i}{2} \,
\int{\frac{\delta}{\delta A_{\mu}} \, \mathbf{D}_{c}^{\mu \nu} \,
\frac{\delta}{\delta A_{\nu}}} \}$-part of the complete linkage
operator appearing in Eq. (\ref{Eq2-6}), and by retaining only the
thermal part of it, that is,
\begin{widetext}
\begin{equation}\label{Eq2-9}
e^{- \frac{i}{2} \int {\frac{\delta}{\delta A_{\mu}}  \delta
\mathbf{D}_{\mu \nu}^{th} \frac{\delta}{\delta A_{\nu}}}} \,
\frac{i}{\tau} \, \sum_{n=-\infty}^{+\infty}{
\int_{0}^{\infty}{ds \, e^{-is(\omega^2-\omega_n^2)} \,
e^{-i\omega_nz_0} \, e^{-ig\int_{0}^{s}{ds' \, \vec{p} \cdot
\vec{A}(y_{0} - 2 s' \omega_{n}, \vec{y} - 2 s' \vec{p})
}}}}\biggr|_{A = 0},
\end{equation}

\noindent which gives,
\begin{equation}\label{Eq2-10}
\frac{i}{\tau} \, \sum_{n=-\infty}^{+\infty}{
\int_{0}^{\infty}{ds \, e^{-is(\omega^2-\omega_n^2)} \,
e^{-i\omega_nz_0} \, e^{2 i g^{2} \, \int_{0}^{s}{ds_{1} \,
\int_{0}^{s}{ds_{2} \, p_{\mu} \delta \mathbf{D}_{\mu
\nu}^{th}{\left((s_1-s_2)p\right)} p_\nu}}}}}.
\end{equation}
\end{widetext}

\noindent In essence, this term's contribution corresponds to the
particle's energy loss due to the bremsstrahlung produced under the
enhancement of the heat bath's photons; for ease of presentation,
that bremsstrahlung produced by the slowing particle inside the heat
bath, will be re-considered in Section \ref{sec3} with the
$\mathbf{D}_{c}^{\mu \nu}$-portion of the linkage operation.

The argument of the last exponential factor of Eq. (\ref{Eq2-10}) can
be written as
\begin{widetext}
\begin{equation}\label{Eq2-11}
2 i g^2 \, \int_{0}^{s}{ds_{1} \, \int_{0}^{s}{ds_{2} \,
\frac{1}{2} \, \int{\frac{d^{3}\vec{k}}{(2\pi)^{3}} \, \frac{1}{k}
\, \frac{e^{2i \vec{k} \cdot \vec{p} \, (s_1-s_2)}}{e^{\beta k}-1} \, 2
\cos[2(s_1-s_2)k\omega_n]\left(\vec{p}^{\, 2}-\frac{({\vec{k}}\cdot{\vec{p}})^2}{{\vec{k}}^{\, 2}} \right)e^{-\frac{k}{p}}}}},
\end{equation}
\end{widetext}

\noindent where a factor $\exp{(-|\vec{k}|/|\vec{p}|)}$ has been inserted as a simple way of enforcing the BN approximation.  Certainly, other such
limitations are possible and the details of the calculation will be
changed somewhat, but the Physics will be essentially the same.
The integrals in Eq. (\ref{Eq2-11}) are well defined, but cannot
be carried out exactly. A sensible, approximate evaluation is proposed in Appendix C, leading to an overall form of
\begin{equation}\label{Eq2-12}
-\xi^{2} g^{2} f(\frac{T}{p})(\vec{p}^{\, 2})^{2} s^{2}, \quad
f(\frac{T}{p}) \simeq \frac{(T/p)^{2}}{1+T/p}
\end{equation}

\noindent where $\xi$ combines some numerical factors and where the
approximation for $f(T/p)$ is valid in the regime $|\vec{p}| \gg T$.
In Eq. (\ref{Eq2-12}), it is worth noticing that power of $s^{2}$;
had we obtained a power of $s$ instead, then we would be lead back
to the scalar field situation in which the whole set of BN
approximations reduced to an exponential temporal damping
of the original free-field result (cf. Eq. (3.35) in Ref. \cite{CFG2005}). As might have been expected in the case of QED, the
$s^{2}$-law leads to a more involved behavior that we now must
evaluate.

One is then left with the expression
\begin{eqnarray}\label{Eq2-13}
& & \nonumber \frac{i}{\tau} \,
\sum_{n=-\infty}^{+\infty}{\int_{0}^{\infty}{ds \,
e^{-is(\omega^2-\omega_n^2)} \, e^{-i\omega_nz_0} \, e^{-a^2s^2}}}, \\ & &
\quad a^2=\xi^{2} \,  g^{2} \, ({\vec{p}}^{\, 2})^{2} \, f(\frac{T}{p})
\end{eqnarray}

\noindent The free-field result, $\tilde{\mathbf{S}}_{th}$ of
Eqs. (\ref{Eq2-3}) and (\ref{Eq2-4}), can still be used to re-write
Eq. (\ref{Eq2-13}) in the form
\begin{widetext}
\begin{equation}\label{Eq2-14}
e^{- a^2 (i \frac{\partial}{\partial \omega^2})^2} \, (2m) \, \frac{i}{\tau} \! \sum_{n=-\infty}^{+\infty}\int_0^\infty{ds \
e^{-is(\omega^2-\omega_n^2)}\ e^{-i\omega_nz_0}=e^{+ a^2
(\frac{\partial}{\partial \omega^2})^2}  \, \tilde{\mathbf{S}}_{th}(\omega,
z_{0})},
\end{equation}
\end{widetext}

\noindent and, by using the representation:
\begin{equation*}
e^{+ a^{2} \, (\frac{\partial}{\partial \omega^{2}})^{2}} =
\frac{1}{\sqrt{\pi}} \, \int_{-\infty}^{+\infty}{db \,
e^{-b^{2} +2 a b (\frac{\partial}{\partial \omega^{2}})}},
\end{equation*}

\noindent it is possible to express our BN-approximated result in
the remarkable and rather simple form of a Gaussian averaged,
$\omega$-translated free-field propagator,
\begin{equation}\label{Eq2-15}
{\mathbf{S}'}_{th}^{BN}(\omega, z_{0}) = \frac{1}{\sqrt{\pi}} \,
\int_{-\infty}^{+\infty}{db \, e^{-b^2} \,
\tilde{\mathbf{S}}_{th}\left({\sqrt{{\omega^{2} - 2ab}}}, z_0\right)}.
\end{equation}

\noindent However simple in principle, an exact integration of
Eq. (\ref{Eq2-15}) remains out of reach. Fortunately, the following
considerations concerning orders of magnitude are helpful in order
to extract the essence of the result.

At small enough coupling constant, one can expect to have $g \xi \leq
1$.  In Appendix \ref{appC}, we found $\xi$ itself is small and less than $1$.  Also, starting from $|\vec{p}| \gg T$, all the way down to thermalization, where $|\vec{p}|$ becomes on the order of $T$, one
has clearly $\sqrt{f} \simeq (T/p)/\sqrt{1+ T/p} < 1$. And finally,
the essential part of Eq. (\ref{Eq2-15}) is given by the range of
$|b| < 1$.  It therefore appears sensible to replace Eq. (\ref{Eq2-15}) by
\begin{equation}\label{Eq2-16}
\frac{1}{\sqrt{\pi}} \, \int_{-\infty}^{+\infty}{db \,
e^{-b^2} \, \tilde{\mathbf{S}}_{th}\left({{{\omega(1+g\xi{\sqrt{f}}b)}}},
z_0\right)}.
\end{equation}

\noindent Inserting Eq. (\ref{Eq2-4}), (with the first, overall
factor of $1/\omega$ left constant for simplicity), integration on
the parameter $b$ can be performed, with a convergent and damped
result of
\begin{widetext}
\begin{equation}\label{Eq2-17}
i \frac{2m}{\omega} \left\{ \frac{1}{2} \, e^{-i \omega z_0} -
e^{-{\omega/ T}+({a/2\omega T})^{2}} \, \cos{\left(\left[\omega - 2
T (\frac{a}{2\omega T})^{2} \right] z_{0}\right)} \right\} \,
e^{-{a^2z_0^2/ 4\omega^2}}.
\end{equation}
\end{widetext}

\noindent As noted in Item 5 of the Introduction, had we retained
only the $n = 0$ term of Eq. (\ref{Eq2-13}), the result would have
had the same structure, but with a slower fall-off in $z_{0}$.  Defining the quantity
\begin{equation}\label{Eq2-18}
q=e^{-{\omega/ T}+({a/2\omega T})^2} \, \cos{\left(\left[\omega - 2T
(\frac{a}{2\omega T})^{2}\right] z_{0}\right)},
\end{equation}

\noindent then, the squared modulus of Eq. (\ref{Eq2-17}) reads
\begin{equation}\label{Eq2-19}
\frac{4 m^{2}}{\omega^{2}} \, \left[ \frac{1}{4}  +  q^2  -  q
\cos{(\omega z_{0})} \right] \, e^{-{a^2z_0^2/ 2\omega^2}}
\end{equation}

\noindent whose leading contribution, in view of Eq. (\ref{Eq2-18}),
is the first term of Eq. (\ref{Eq2-19}),
\begin{equation}\label{Eq2-20}
\frac{m^{2}}{\omega^{2}} \, \exp{\left(-\frac{z_{0}^{2} \alpha
T^{2}}{1+T/p}\right)} \rightarrow \frac{m^{2}}{\omega^{2}} \,
\exp{\left(- z_{0}^{2} \alpha T^{2} \right)},
\end{equation}

\noindent the meaning(s) of which will be discussed in detail in Section \ref{sec3}.  In order to represent the thermalization of such an incident particle, one may introduce the ratio of Eq. (\ref{Eq2-20}), taken at
a given $z_{0}$-value, to its initial value at $z_{0} = 0$.

Let $R(z_{0} T, p(z_{0})/T, \alpha) \equiv R(z_{0})$ be this ratio.
One has
\begin{equation}\label{Eq2-21}
R(z_{0})= \left( \frac{p(0)}{p(z_{0})} \right)^{2} \, \exp{\left(
-\frac{z_{0}^{2} \, \alpha T^{2}}{1+T/p(z_0)} \right)},
\end{equation}

\noindent where we have resorted to the approximation of $\omega \simeq p$, which agrees with the ordering of scales: $p \gg T \gg m$.  Restoring all of the conventional units, one can write
\begin{equation}\label{Eq2-22}
z_{0} T \rightarrow \frac{z_{0} k_{B} T}{\hbar}=\frac{c
z_{0}}{\lambda_{c}} \, \frac{k_{B} T}{m c^{2}},
\end{equation}

\noindent where $k_{B}$ is the Boltzmann constant and $\lambda_{c}$
the Compton wavelength of the traveling particle.  We assume a reasonable, semi-classical model for the decrease of
$\vec{p} = \vec{p}(z_0)$, of form $\vec{p}(0) \exp{(-\Gamma z_{0})}$
with $\Gamma = \Gamma_{\mathrm{Doppler}} = \xi \alpha c/\lambda_{c} \,
(kT/mc^{2})^{2}$, where $\xi$ is a numerical constant and
$\lambda_{c} =\hbar/mc$, as derived in Appendix \ref{appB}.
Then,
\begin{equation}\label{Eq2-23}
p(z_{0})=p(0) \, e^{-\Gamma z_{0}}, \quad \Gamma = \xi \frac{\alpha
c}{\lambda_{c}} \, \left( \frac{k_{B} T}{m c^{2}} \right)^{2},
\end{equation}

\noindent and the rise and subsequent fall-off of $R(z_{0})$ can simply
be read off from the expression
\begin{equation}\label{Eq2-24}
R(z_{0})= \exp{\left\{ \left( \frac{k_{B} T}{m c^{2}} \right)^{2} \,
\frac{c z_{0}}{\lambda_{c}} \, \left[ 2 \xi \alpha - \frac{c
z_{0}}{\lambda_{c}} \frac{1}{1+T/p(z_{0})} \right] \right\} }.
\end{equation}

\noindent If one assumes a specific form for the "linear" density of
equilibrium heat bath photons in the Doppler computation, e.g.,
$\rho(\nu)= N(\nu) = [\exp{(h \nu / k_{B} T)} - 1]^{-1}$, that is,
the conventional Planck distribution, then the constant $\xi$ can be
evaluated, and a specific value of $z_{0}$ predicted for when the
exponential factor of Eq. (\ref{Eq2-24}) vanishes, and the fireball
starts to decrease.

One finds, for example,
\begin{equation}\label{Eq2-25}
z_{0} =2 \xi \alpha \frac{\lambda_{c}}{c} - \frac{1}{\Gamma} \,
W_{0}{\left( - 2\Gamma \xi \alpha \, \frac{\lambda_c}{c} \, \frac{T}{p_0} \, e^{2 \left( \frac{\xi \alpha k_{B} T}{m c^{2}} \right)^{2}}
\right)}
\end{equation}

\noindent as the time after which a possible fireball
starts to decrease. In Eq. (\ref{Eq2-25}), $W_{0}$ is the principal
branch of the Lambert-W-function \cite{Lambert}. Provided its
argument lies within the convergence radius of $1/e$, one has
\begin{equation}\label{Eq2-26}
2 (\frac{T}{p_0}) \, \left(\frac{\xi \alpha k_{B} T}{m c^{2}}
\right)^{2} \, \exp{\left[ {2\left(\frac{\xi \alpha k_{B} T}{m
c^{2}} \right)^{2}} \right]} \leq e^{-1},
\end{equation}

\noindent and then, $W_{0}$ can be replaced by its series expansion, and Eq. (\ref{Eq2-25}) may be approximated as
\begin{equation}\label{Eq2-27}
z_{0} \simeq 2 \xi \alpha \frac{\lambda_c}{c} \, \left( 1+
\exp{\left[ 2 \left(\frac{\xi \alpha k_{B} T}{m c^{2}} \right)^{2}
\right]} + \cdots \right).
\end{equation}

\noindent Note that $\mathcal{O}(\xi \alpha k_{B} T/mc^{2}) <1$ is a necessary condition for Eq. (\ref{Eq2-26}) to be satisfied, and the series expansion of $W_{0}$ to be reliable.  In a system of natural units such as $\hbar=c=k_{B}=1$, this is equivalent to ${\cal{O}}(\alpha T/m) <1$, and this condition somewhat specifies, and restricts, the amount by which $T$ is assumed to be much larger than $m$ (remember the assumed ordering of $p \gg T \gg M$).  One has then for $\Delta z_{0}$ an estimation on the order of $\alpha/m$.

\section{\label{sec3}Fermionic loops}

In order to include pair-production as a mechanism for the loss of
the incident particle's energy, it is necessary to include at least
the simplest closed-lepton-loop, whose absorptive part corresponds
to the probability of pair production. Let us approximate $\mathbf{L}[A]$,
defined in Appendix \ref{appA}, in the simplest way, as $\mathbf{L}[A] =
\frac{i}{2} \int{ \int{ A_{\mu}(x) \mathbf{K}^{\mu \nu}(x-y) A_{\nu}(y)}}$,
where the gauge invariant representation of $\mathbf{K}_{\mu\nu}$ reads \cite{HMF1972}
\begin{eqnarray}\label{Eq3-1}
\tilde{\mathbf{K}}_{\mu\nu}(k) &=& - k^{2} (\delta_{\mu\nu} - \frac{k_{\mu} k_{\nu}}{k^{2}}) \, \Pi(k^2) \\ \nonumber &=& - k^{2} (\delta_{\mu\nu} - \frac{k_{\mu} k_{\nu}}{k^{2}}) \, [\Pi(0) + \Pi_{R}(k^{2})]
\end{eqnarray}

\noindent with the renormalized part of $\Pi_{R}$,
\begin{equation}\label{Eq3-2}
\Pi_{R}(k^2) = - \frac{2 \alpha}{\pi} \int_{0}^{1}{du \, u(1-u)
\ln{\left(1+u(1-u)\frac{k^{2}}{m^{2}}\right)}}
\end{equation}

\noindent Here, $m$ is the mass of the looping fermions and
renormalization has been performed so that $\Pi_{R}(0) = 0$.  It is
a real-valued function of $k^2$, as long as $k_0^2 < \vec{k}^{\, 2} + 4
m^2$, but if $k_0^{2} > \vec{k}^{\, 2} + 4 m^{2}$, it develops an imaginary
part given by the discontinuity of the logarithm across the cut, of
value $2i\pi \Theta(k_{0} - \sqrt{{\vec{k}}^{\, 2} +4 m^{2}})$. The real part of the
logarithm, for large $\vec{k}^{\, 2}$, is proportional to $\ln(\vec{k}^{\, 2}
/ m^2)$, plus additive constants.

We have now, instead of Eq. (\ref{Eq2-6}), the expression
\begin{widetext}
\begin{equation}\label{Eq3-3}
{\langle \vec{p}, n| {\mathbf{S}'}^{BN}_{th} |\vec{y}, y_{0} \rangle} = \left.
e^{- \frac{i}{2} \int{\frac{\delta}{\delta A_{\mu}}  {D}_{th}^{\mu
\nu} \frac{\delta}{\delta A_{\nu}}}} \, {\langle \vec{p}, n|
\mathbf{G}_{th}[A] |\vec{y}, y_{0} \rangle} \, \frac{e^{\mathbf{L}[A]}}{\mathbf{Z}[i\tau]} \right|_{A = 0},
\end{equation}
\end{widetext}

\noindent where $\mathbf{Z}[i\tau]$ is the normalization factor corresponding to the interacting partition function.

Using the BN approximation specified in Eq. (\ref{Eq2-5}),
as well as the simplest approximation to $\mathbf{L}[A]$, the functional
differentiation of Eq. (\ref{Eq3-3}) can be performed exactly with
the help of the functional identity
\begin{eqnarray}\label{Eq3-4}
& & \exp{\left[- \frac{i}{2} \int{\int{\frac{\delta}{\delta
A} \cdot \mathbf{D}_{th} \cdot \frac{\delta}{\delta A}}}\right]} \\ \nonumber & & \quad \quad \cdot  \left.  \exp{\left[\frac{i}{2} \, \int{ \int{ A \cdot \mathbf{K} \cdot A}} - i \int{ f \cdot A}\right]} \right|_{A = 0} \\ \nonumber &=& \exp{ \left[ \frac{i}{2} \int{ \int{f \cdot \mathbf{D}_{th} \frac{1}{1 - \mathbf{K} \cdot \mathbf{D}_{th}} \cdot f}} \right]} \\ \nonumber & & \quad \quad \cdot \exp{ \left[-\frac{1}{2} \mathrm{Tr}{\ln{(1 - \mathbf{K} \cdot \mathbf{D}_{th})}} \right]},
\end{eqnarray}

\noindent where $f_{\mu}= g p_{\mu} \int_{0}^{s}{ds' \, \delta(x-(y-s'p))}$ and
the Trace-Log determinantal factor of Eq. (\ref{Eq3-4}) has no
relation to the traveling particle and is absorbed into the
partition-function relation
\begin{equation}\label{Eq3-5}
\mathbf{Z}[\beta] = \left. e^{-\frac{1}{2} \mathrm{Tr}{\ln{(1 - \mathbf{K} \cdot
\mathbf{D}_{th})}}} \right|_{\tau \rightarrow -i \beta} \cdot \mathbf{Z}_{0}[\beta].
\end{equation}

\noindent The resulting thermal propagator has the same form as
given in Eqs. (\ref{Eq2-9}) and (\ref{Eq2-10}), except that the term
$\exp{\{2 i g^{2} \int{\int{p \cdot \delta\mathbf{D}_{th} \cdot p}}\}}$ is now replaced by
\begin{widetext}
\begin{equation}\label{Eq3-6}
\exp{\left\{ 2 i g^{2} \, \int_{0}^{s}{ ds_{1} \, \int_{0}^{s}{
ds_{2} \int{{\frac{d^{4}k}{(2\pi)^4} \, \left[ p \cdot \mathbf{D}_{th} \left(\frac{1}{1 - \mathbf{K} \cdot \mathbf{D}_{th}}\right)  \cdot p \right] \, e^{-2 i (s_1 - s_2)(k_0 \omega_n - \vec{k} \cdot \vec{p} )}} }}} \right\}}.
\end{equation}
\end{widetext}

\noindent Recalling $\mathbf{D}_{th}^{\mu \nu}= \mathbf{D}_{c}^{\mu \nu} + \delta
\mathbf{D}_{th}^{\mu \nu}$, Eq. (\ref{Eq2-7}), and $k^{2} \, \mathbf{D}_c(k) = \mathbf{\hat{D}}$, the
denominator in the large parenthesis of Eq. (\ref{Eq3-6}), $1 - \mathbf{K}
\cdot \mathbf{D}_{th}$, reduces to $1 + \Pi(k^{2})$, whereas the numerator,
with $\mathbf{D}_{th}(k)$, separates into two distinct parts, of which we
consider first the contribution coming from $\mathbf{D}_{c}(k)$
\begin{eqnarray}\label{Eq3-7}
& & \exp{\left[\frac{i}{2} \int{f \cdot \mathbf{D}_{c} \, \frac{1}{1 +
\Pi(k^{2})} \cdot f} \right]}  \\ \nonumber &=& \exp{ \left[-
\frac{i}{2} \int{ f \cdot \mathbf{D}_{c} \, \frac{1}{1 + \Pi(0)} \,
\frac{1}{1+ \frac{\Pi_{R} }{1+ \Pi(0)}} \cdot f } \right]} \\
\nonumber &\rightarrow& \exp{\left[\frac{i}{2} \int{ f \cdot \mathbf{D}_{c}
\cdot f} \right]} \cdot \exp{ \left[- \frac{i}{2} \int{ f \cdot \mathbf{D}_{c}
\, \frac{\Pi_{R} }{1+ \Pi_{R}} \cdot f } \right]},
\end{eqnarray}

\noindent where the factor of $[1 + \Pi(0)]^{-1}$ renormalizes all $g^{2}$-dependence in the last line, since
$Z_{3}^{-1} = 1 + \Pi(0)$ and $g_{R}^{2} = Z_{3} \, g^{2}$.

We first consider the first exponential factor on the right hand side of
Eq. (\ref{Eq3-7}): In a $T = 0$, non-BN calculation where $k
> p$ is permitted, this term generates the UV divergences associated
with mass and wave-function renormalization, and those terms should
properly be discarded as in Section \ref{sec2}.  In a $T > 0$
context, this term describes the damping of the particle's energy
due to ordinary bremsstrahlung (in contrast, the remaining part related to $\delta \mathbf{D}_{th}$ in
Eq. (\ref{Eq3-6}) describes the damping of the particle's energy as
"enhanced" by the thermal photon heat-bath in which the particle is slowing
down).

The evaluation of this first term begins with
\begin{widetext}
\begin{equation}\label{Eq3-8}
\exp{\left\{4 i g^{2} p_{i} p_{j} \, \int_{0}^{s}{ds_{1} \,
\int_{0}^{s_{1}}{ds' \, \int{ \frac{d^{4}k}{(2\pi)^{4}} \,
\frac{e^{-2is'(k_0 \omega_n - {\vec{k}} \cdot \vec{p})}}{k^{2} - i
\varepsilon} \, (\delta_{ij} - \frac{k_i k_j}{\vec{k}^2}) \,
e^{-k/p} }}} \right\}},
\end{equation}
\end{widetext}

\noindent and the integration over $k_0$ can be carried out by
contour integration:
\begin{eqnarray}\label{Eq3-9}
& & - \int{dk_{0}\frac{e^{-2i s' k_0 \omega_n}}{[k_0 - (k - i
\varepsilon)][k_0 + (k - i \varepsilon)]}} \\ \nonumber &=& + \frac{i \pi}{k}
\left[ \Theta(n) e^{-2is'k\omega_n} + \Theta(-n) e^{2is'k\omega_n}
\right].
\end{eqnarray}

\noindent Relying again on the approximations used in Section
\ref{sec2}, which amount basically to the neglect of the oscillating
factors of Eqs. (\ref{Eq3-8}) and (\ref{Eq3-9}), Eq. (\ref{Eq3-9})
then becomes simply the quantity $i \pi/k$. Inserted into
Eq. (\ref{Eq3-8}), one gets immediately a contribution of
\begin{equation}\label{Eq3-10}
\exp{\left[ - \frac{4}{3 \pi} \alpha ({\vec{p}}^{\, 2})^{2} s^{2} \right]}.
\end{equation}

\noindent That is, the first term in the right hand side of
Eq. (\ref{Eq3-7}) adds the amount $\frac{4}{3 \pi} \alpha
(\vec{p}^{\, 2})^{2}$ to the $a^{2}$-constant which appears in
Eq. (\ref{Eq2-13}), an additional damping independent of the
thermalized heat bath (at this level of approximation, of course).

The remaining factor of Eq. (\ref{Eq3-7}) displays a nice example of
the basic unitarity of QED: even though we have used the lowest
$g^{2}$-order approximation to $\mathbf{L}[A]$, in conjunction with our BN
treatment, a very large $\Pi_{R}$ cannot produce an overly large
effect, for automatic damping (in the Hartree-Fock sense) is
provided by its denominator. At first, we shall assume a weak
effect, and accordingly replace that denominator by 1; and then,
subsequently, the modifications will be noted when the complete
denominator is used.  But before we proceed with that very contribution, an interesting point must be made concerning the contributions attached to the $\delta \mathbf{D}_{th}^{\mu \nu}$-piece of the full $\mathbf{D}_{th}^{\mu \nu}$ propagator appearing in Eq. (\ref{Eq3-6}).

\par The contribution attached to $\delta \mathbf{D}_{th}^{\mu \nu}$ in
Eq. (\ref{Eq3-6}) may be written in a way similar to Eq. (\ref{Eq3-7})
\begin{eqnarray}\label{Eq3-11}
& & \exp{\left[ \frac{i}{2} \int{f \cdot \delta \mathbf{D}_{th} \frac{1}{1 +
\Pi(k^{2})} \cdot f} \right]} \\ \nonumber &=& \exp{\left[\frac{i}{2} \int{f \cdot \frac{\delta \mathbf{D}_{th}}{1+\Pi(0)} \cdot f} \right]} \\ \nonumber & & \quad \quad \cdot \exp{\left[- \frac{i}{2} \int{f \cdot \frac{\delta \mathbf{D}_{th}}{1+\Pi(0)} \frac{\Pi_R}{ 1 + \Pi(0)} \cdot f}\right]},
\end{eqnarray}

\noindent where the first factor, in the right hand side of
Eq. (\ref{Eq3-11}), is that part already calculated in Section
\ref{sec2} which leads to the result of Eq. (\ref{Eq2-13}). The second
term would correspond to a pair production mechanism, enhanced by
the thermal heat bath photons. However, since $\Pi_R(k^2)$ vanishes
at $k^2 = 0$, this term vanishes because $\delta \mathbf{D}_{th}(k)$ is
proportional to $\delta(k^2)$ (see Eq. (\ref{Eq2-7})).  Over the relevant $k$-integration range, the leading thermal contribution to $\Pi(k)$, that is the so-called {\it{Hard Thermal Loop}} (HTL) polarization tensor, $\Pi^{HTL}(k_0, {{k}})$, can be as large and even larger than the renormalized T=0-part, $\Pi_R(k^2)$, \cite{TGPR1997,CG2005a}.  However, this leading thermal piece of $\Pi^{HTL}(k_0, {{k}})$, is also proportional to $k^2$ and therefore does not contribute either, to the second term on the right hand side of Eq. (\ref{Eq3-11}).

\par
Returning to the second right hand side factor of Eq. (\ref{Eq3-7})
in its lowest $g^{4}$-order, $\exp{\{-(i/2) \int{f \cdot \mathbf{D}_c \Pi_R \cdot f}\}}$,
one needs to evaluate
\begin{widetext}
\begin{equation}\label{Eq3-12}
\exp{\left\{ 4 i g^2 \int_{0}^{s}{ds_{1} \,
\int_{0}^{s_1}{ds' \, \int{\frac{d^{4}k}{(2\pi)^4} \,
\frac{e^{-2is'(k_0 \omega_n - \vec{k} \cdot \vec{p})}}{k^2 - i
\varepsilon} \, \Pi_{R}(k^2) \, \left(\vec{p}^{\, 2} -
\frac{(\vec{p} \cdot \vec{k})^{2}}{\vec{k}^{\, 2}}\right) \, e^{-k/p}
}}} \right\}},
\end{equation}
\end{widetext}

\noindent and focus interest on the contribution coming from the
imaginary part of $\Pi_R(k^2)$, given by $-(2 i \alpha/3)\, \Theta
(k^2 - 4 m^2)$.  Note that the renormalization prescription of
$\Pi_R(0)=0$ ensures that there is no singularity at $k^{2} =0$.  And,
therefore, the $k_{0}$-integral over the real part of
$\Pi_{R}(k^{2})$ receives no contribution \cite{Sheu2008a}.

To evaluate the absorptive part of Eq. (\ref{Eq3-12}), consider first
\begin{eqnarray}\label{Eq3-13}
& & \int{dk_0 \, \frac{\Theta(k_0 - \sqrt{\vec{k}^{\, 2} + 4 m^2})}{[k_0-k+i\varepsilon][k_0+k-i\varepsilon]}}
\\ \nonumber &\rightarrow& \frac{1}{2k} \int_{\sqrt{\vec{k}^{\, 2} + 4 m^2}}^{\Lambda}{dk_0 \, (\frac{1}{k_{0} - k} - \frac{1}{k_{0} + k})},
\end{eqnarray}

\noindent where the upper limit of $\Lambda$ cannot be chosen larger
than the particle's available energy (which it gives to the virtual
photon, which then produces the pair).  And even though one cannot
make a mass-shell measurement of the particle as it passes through
the medium, its energy surely cannot be too far from its mass-shell
value, which is essentially $cp$ (until thermalization occurs, $cp >
k_{B} T$). Taking $\Lambda$ on the order of $p$, the oscillating
factors of Eq. (\ref{Eq3-12}) are again sufficiently small to be
neglected, and what remains is the simple integral
\begin{eqnarray}\label{Eq3-14}
& & -\frac{1}{2k} \int_{\sqrt{\vec{k}^{\, 2} + 4m^2}}^{p}{dk_{0} \,
(\frac{1}{k_0 - k} - \frac{1}{k_0 + k})} \\ \nonumber &=& \frac{1}{k} \left\{
\ln{\frac{p + k}{p - k}} + \ln{\frac{\sqrt{\vec{k}^{\, 2} + 4 m^2} - k
}{\sqrt{\vec{k}^{\, 2} + 4 m^2} + k}} \right\}.
\end{eqnarray}

\noindent With $k = xp$, this combination can be approximately reduced to
\begin{equation}\label{Eq3-15}
- \frac{1}{k} \left[ \ln{(1-x^2)} + \ln{(\frac{\vec{p}^{\, 2}}{4 m^2})}
\right].
\end{equation}

Then, combining all factors and retaining only the most important
$\ln(\vec{p}^{\, 2}/m^{2})$ dependence, one finds for the absorptive
contribution of Eq. (\ref{Eq3-12}), the amount
\begin{equation}\label{Eq3-16}
-\frac{4}{3\pi} \, {\left(\frac{g^2}{4\pi}\right)^{2}} \, s^2
(\vec{p}^{\, 2})^2 \ln{\left(\frac{\vec{p}^{\, 2}}{m^2}\right)}.
\end{equation}

\noindent The remaining denominator factor of Eq. (\ref{Eq3-7}) can
be taken into account by writing $1 + \Pi_R(k^2)= 1 + g^{2}(u -
iv)$, and identifying the new absorptive part of $\Pi_R /[1 +
\Pi_R]$ as $-i g^{2} v/[(1 + g^{2} u)^{2} + g^{4} v^{2}]$. Here, one
has $u = \mathcal{O}(\ln(\vec{k}^{\, 2}/m^2))$, which, after integration,
translates into a denominator factor of $\ln(\vec{p}^{\, 2}/m^2)$, thereby
removing the $\ln(\vec{p}^{\, 2}/m^2)$ factor of Eq. (\ref{Eq3-16}), and
effectively substituting a factor of $[\ln(\vec{p}^{\, 2}/m^{2})]^{-1}$.
However, other, higher-order corrections from the photon polarization
may change the result.  From this simple photon bubble, in our BN
approximation and for small coupling, there appears little
pair-production enhancement of the Bremsstrahlung damping of
Eq. (\ref{Eq3-10}).

In our calculation, pair-production is considered as one of processes of energy depletion of the incident particle.  The fermion-loop pairs are not considered as thermalized at the instant of production, and have no knowledge of the medium without subsequent interaction, which is irrelevant to the incident particle's energy loss.  In contrast, previous calculations \cite{BI1996a,BI1997a,BI1997b} have replaced internal photon lines with effective (resummed) photon propagators in the HTL approximation, in which the closed-fermion-loop momenta are assumed to be larger than those of soft thermal photons in the construction of the photon polarization tensor \cite{Braaten1990a, Braaten1990b,Braaten1990c,Frenkel1990a}.  The loop fermions we use are not taken as thermalized; rather, we employ the conventional, renormalized, photon polarization tensor, and extract its imaginary contribution as that piece of the calculation relevant to pair production.

\section{\label{sec4}Transverse vs. Longitudinal Fireballs}

This Section develops the interpretation of the resulting fermionic two-point function which has been introduced in Ref. \cite{CFG2005}. Collecting all three damping factors, with $p \gg T$, the $a^{2}$-constant in
Eq. (\ref{Eq2-13}) now becomes
\begin{eqnarray}\label{Eq4-1}
\nonumber \tilde{a}^{2} &=& 4 \pi \xi^{2} \alpha T^{2} \vec{p}^{\, 2} + \frac{4}{3 \pi}
\alpha ({\vec{p}}^{\, 2})^2 - \frac{4}{3\pi} \alpha^{2} (\vec{p}^{\, 2})^2
\ln{\left(\frac{\vec{p}^{\, 2}}{m^2}\right)} \\ &=& \frac{4}{3\pi} \alpha (\vec{p}^{\, 2})^{2} \left\{ \left[1 + \left(\frac{2\pi T}{p}\right)^{2}\right]- \alpha \ln{\left( \frac{\vec{p}^{\, 2}}{m^{2}}\right)} \right\},
\end{eqnarray}

\noindent where it is encouraging to recognize a term of $1 + (\frac{2 \pi T}{p})^{2}$, peculiar to rigorous one-loop perturbative calculations \cite{TGPR1997}, and Eq. (\ref{Eq2-17}) can be re-written as
\begin{widetext}
\begin{equation}\label{Eq4-2}
i \frac{2m}{\omega} \left\{ \frac{1}{2} \, e^{-i \omega z_0
-\frac{\tilde{a}^2}{4 \omega^2} z_{0}^{2}} - e^{-\frac{\omega}{T} -
\frac{\tilde{a}^2}{4 \omega^2}(z_{0}^{2} - \frac{1}{T^{2}})} \,
\cos{\left(\left[\omega - 2 T (\frac{\tilde{a}}{2\omega T})^{2}\right]
z_{0}\right)} \right\}.
\end{equation}
\end{widetext}

\noindent The second term of Eq. (\ref{Eq4-2}) describes the
disturbance inside the heat bath which is isotropic in the medium.
Were one to calculate the spatial thermal propagator, by a Fourier
transform over $\vec{p}$ at any given time (before thermalization) in the
medium, the cosine factor of Eq. (\ref{Eq4-2}) would correspond to the
appearance of a disturbance propagating with exponential phase
factors of $[i(\vec{p} \cdot \vec{z} - Q z_{0})]$ and $[i(\vec{p} \cdot \vec{z} + Q z_{0})]$,
with $Q$ the square-bracket constant that appears in the cosine's argument of Eq. (\ref{Eq4-2}); and if the
dummy variable $\vec{p}$ is changed to $-\vec{p}$ in the integration over the
second exponential factor, the result suggests the propagation of
symmetric, "transverse" back-to-back pulses in any arbitrary
direction.  This is true for the free propagator and interacting propagator,
and is to be expected of a thermal Green's function
which not only describes the effects of an incident particle
entering the medium, but also contains a description of any
"tsunami-like" disturbance originating in the medium. [One may think
here of the emission by the incident particle of a virtual photon
with high-energy and little momentum, which immediately decays into
an electron-positron pair, which then comprise and sequentially
generate the corresponding transverse fireballs.]

The factor, $e^{-\omega/T}$, in the second term of Eq. (\ref{Eq4-2}) comes from the Fermi-Dirac distribution
and the combined factor, $\frac{1}{\omega} \, e^{-\frac{\omega}{T} +
\frac{\tilde{a}^2}{4 \omega^{2} T^{2} }}$, determines the initial relative
magnitude of this second term compared to the first.  The phase is also different from that of the incoming fermion by a shift of $2 T (\frac{\tilde{a}}{2\omega T})^{2}$.  The square modulus of this second term governs what we have called the "transverse" fireball, as defined in the Item 6 of the Introduction,
\begin{equation}\label{Eq4-3}
\frac{4 m^{2}}{\omega^{2}} \, e^{-\frac{2\omega}{T} - \frac{\tilde{a}^2}{2
\omega^2}(z_{0}^{2} - \frac{1}{T^{2}})} \,
\cos^{2}{\left(\left[\omega - 2 T (\frac{\tilde{a}}{2 \omega T})^{2}\right]
z_{0}\right)}.
\end{equation}

\noindent Similar to the analysis in Section \ref{sec2} for the
longitudinal counterpart which now reads
\begin{equation}\label{Eq4-4}
\frac{ m^{2}}{\omega^{2}} \, \exp{\left[-\frac{\tilde{a}^{2} z_{0}^{2}}{2 \omega^{2}}\right]}.
\end{equation}

\noindent One can see in Eq. (\ref{Eq4-3}) that the magnitude of the transverse
disturbance rises as the incoming fermion starts to lose
momentum/energy as $1/\omega^{2}$.  However, the exponent decrease
from its initial value is Gaussian and much faster than
$1/\omega^{2}$ as time goes on.  Then the fireball shrinks in all
directions once $z_{0}^{2} > T^{-2}$ (or $z_{0}^{2} \hbar^{-2} >
(k_{B} T)^{-2}$).  Hence, a very simple prediction arises from this BN-approximated QED calculation; which is a relative increase of the transverse disturbance as $z_0$ increases, over a duration extent of $\Delta z_0=1/T$.

As compared to Eqs. (\ref{Eq2-26}) and (\ref{Eq2-27}) for the longitudinal excitation (and
neglecting the modification due to $a^{2} \rightarrow \tilde{a}^{2}$), one
would therefore have, because of $m > \alpha T$, the inequality $\Delta
z_{0}^{(T)} > \Delta z_{0}^{(L)}$ for the durations after which
transverse $(T)$ and longitudinal $(L)$ excitations quickly decay.

\section{\label{sec5}Summary}

Here, then, are all the damping factors estimated in a strict Bloch-Nordsieck (BN)
framework, where there are no ultraviolet divergences and no
infrared divergences. The evaluations involving the neglect of
weakly-oscillating integrands are of course approximate, but quite
reasonable; and the results are physically correct in the sense that
three sources of momentum and energy loss are included.  It is worth noting that the results
are also elegant in that they involve the complete thermal
propagator and provide a continuous time-dependence of the process,
a "fireball" growth, followed by thermal decay which is Gaussian,
rather than a simple exponential.  We emphasize that, using straight-forward functional methods, we sum over relevant contributions of the thermalized photons in the process of calculating the rapid thermalization of the incident particle.  In contrast to the HTL approach, we do not consider fermion-loop lines which define pair-production to be thermalized.

In this article, the case of a rapid massive fermion entering a
thermalized QED plasma has been considered.  Through the evaluation of the thermal fermion propagator, our
focus has been to investigate depletion mechanisms that, in the
ideal case of an infinite plasma, bring the incident fermion down to thermal equilibrium.  The same formalism generates the probability requirement for the appearance of shock waves developing in the thermalized medium \cite{CFG2005}.

In order to go beyond the limitations of pure one-loop perturbative
calculations, the present estimate is carried out in a non-
perturbative way with a BN formalism associated with a realistically-decreasing incident particle momentum. Not only do these approximations open the road to
tractable calculations, but hopefully, they should also be physically
relevant to the processes under consideration.

Within a convenient real-time/imaginary-temperature formalism,
calculations are first carried out with the help of the quenching
approximation, and we find a simple and elegant expression in terms
of the free, non-interacting thermal fermion propagator.  Then, the quenching approximation
is relaxed by taking leading effects of fermion loops into account. And
remarkably enough, up to the redefinition of a key parameter, the
simple form of the "quenched result" is preserved.

For the incident particle, energy depletion mechanisms are taken
to be induced by bremsstrahlung and pair production, whereas an intuitive, semi-
classical Doppler model is formulated to account for
momentum damping.  In contrast to previous calculations a complete
time-dependent description of the physical processes at play is
obtained.  In particular, the possibility of longitudinal and transverse shock waves is
seen to develop in the plasma with different amplitudes and phases.
Over two different scales of time duration, both excitations start to
increase, and then quickly decay with a gaussian law.  This relatively simple QED analysis was motivated by the experimental runs at RHIC; and it will be interesting to learn, in a future investigation, if these relatively simple results also hold in QCD.

\begin{acknowledgments}
H.M. Fried would like to thank the J. Schwinger Foundation for a travel grant JSF07080000 which contributed to the research and writing of this paper.
\end{acknowledgments}

\appendix

\section{\label{appA}On the functional formalism}

The fully dressed Thermal Green's function is
\begin{equation}\label{EqA-1}
\mathbf{S}'_{th} = \left. {e^{\mathfrak{D}^{th}_{A}} \cdot \left[ \mathbf{G}_{th}[A] \,
\frac{e^{\mathbf{L}_{th}[A]}}{\mathbf{Z}[i \tau]} \right]} \right|_{A \rightarrow 0},
\end{equation}

\noindent where $\mathbf{L}_{th}[A] = \mathrm{Tr} \ln{[1 - i g (\gamma \cdot A) \,
\mathbf{S}_{th}]}$ and the $\mathfrak{D}^{th}_{A}$ operator of the linkage in the configuration space representation is
\begin{equation}\label{EqA-2}
\mathfrak{D}^{th}_A = - \frac{i}{2} \int{dx \int{dy \, \frac{\delta}{\delta A_{\mu}(x)} \cdot \mathbf{D}_{th}^{\mu \nu}(x-y) \cdot \frac{\delta}{\delta A_{\nu}(y)}}}.
\end{equation}

\noindent  The thermal fermion Green's function $\mathbf{G}_{th}[A]$ is taken in the conventional Matsubara formalism but a Matsubara representation is not needed and is not used for the thermal photon propagator of the linkage operator. In the momentum representation of our formalism, the thermal photon propagator is separated into two parts,
\begin{equation}\label{EqA-3}
\mathbf{D}^{\mu \nu}_{th} = \mathbf{D}^{\mu \nu}_{c} + \delta \mathbf{D}^{\mu \nu}_{th}
\end{equation}

\noindent with a corresponding splitting of the linkage operator Eq. (\ref{EqA-2}). The linkage operation can accordingly separate into two steps, the causal ($T=0$) and thermal ($T \neq 0$) part, and the order of functional operation can be exchanged. For example,
\begin{equation}\label{EqA-4}
\mathbf{S}'_{th} = \left. {\left\{ e^{\Delta \mathfrak{D}^{(th)}_{A}} \, \left[
e^{\mathfrak{D}^{(c)}_{A}} \left( \mathbf{G}_{th}[A] \,
\frac{e^{\mathbf{L}_{th}[A]}}{\mathbf{Z}[i \tau]} \right) \right] \right\}}
\right|_{A \rightarrow 0}
\end{equation}

In addition to the Bremsstrahlung effect at $T>0$, the first linkage operation, with $\exp[{\mathfrak{D}}^{(c)}_{A}]$, will produce factors of mass and wave-function renormalization, so that
Eq. (\ref{EqA-4}) may be written approximately as
\begin{equation}\label{EqA-5}
{\mathbf{S}'}_{th} = \left.{\left\{ e^{\Delta \mathfrak{D}^{(th)}_{A}} \left(
\mathbf{G}_{th,R}[A] \, {\frac{e^{\mathbf{L}_{th,R}[A]}}{\mathbf{Z}_{,R}[i\tau]}} \right)
\right\} } \right|_{A \rightarrow 0},
\end{equation}

\noindent where mass and wave function renormalizations that have nothing to do with the medium are included
in the fermion Green's functional, $\mathbf{G}_{th,R}[A]$, and closed-fermion-loop
functional, $\mathbf{L}_{th,R}[A]$.  For notational simplicity, the
renormalization symbol, $R$, will be dropped in the following, and the mixed representation will hold
\begin{eqnarray}\label{EqA-6}
& & {\langle \vec{p}, n| {\mathbf{S}'}_{th} |\vec{y}, y_{0} \rangle} \\ \nonumber &=&  \left.{
e^{\Delta \mathfrak{D}^{(th)}_{A}}  \left( {\langle \vec{p}, n| \mathbf{G}_{th}[A]
|\vec{y}, y_{0} \rangle} \, {\frac{e^{\mathbf{L}_{th}[A]}}{\mathbf{Z}[i \tau]}}
\right) }\right|_{A \rightarrow 0},
\end{eqnarray}

\noindent where the "quenching" approximation is used; that is, the
fermion determinant is suppressed, and Eq. (\ref{EqA-6}) is replaced (with the subscript $Q$ for
"Quenched") by
\begin{eqnarray}\label{EqA-7}
& & {\langle \vec{p}, n| {\mathbf{S}'}_{th} |\vec{y}, y_{0} \rangle}_{Q} \\ \nonumber &=& \left. \frac{1}{\mathbf{Z}_{0}[i \tau]} \, e^{\Delta \mathfrak{D}^{(th)}_{A}} {\langle
\vec{p}, n| \mathbf{G}_{th}[A] |\vec{y}, y_{0} \rangle} \right|_{A
\rightarrow 0},
\end{eqnarray}

\noindent where $\mathbf{Z}_{0}[i \tau]$ is the free partition function whose relation to $ \mathbf{Z}[i \tau]$ is the following,
\begin{equation}\label{EqA-7b}
\mathbf{Z}[i \tau] = \left. e^{\mathfrak{D}^{(th)}_{A}} \, e^{\mathbf{L}_{th}[A]} \right|_{A
\rightarrow 0} \cdot \mathbf{Z}_{0}[i \tau]. \nonumber
\end{equation}

\noindent One finds \cite{HMF2002,Sheu2008a}
\begin{widetext}
\begin{eqnarray}\label{EqA-8}
& & { \langle \vec{p}, n| {\mathbf{S}'}_{th} |\vec{y}, y_{0}
\rangle}_{Q} \\ \nonumber &=& (\mathbf{Z}_{0}[i \tau])^{-1}
\left[(2\pi)^{3} \tau \right]^{-1/2} e^{-i(\vec{p}\cdot\vec{y} -
\omega_{n}y_{0})}\ i \int_{0}^{\infty} ds
\ e^{-is (m^2 + p^2)} e^{- \frac{1}{2} \mathrm{Tr}{\ln\left(2h\right)} }\\
\nonumber  & & \quad \times \int{d[w]\ \exp{\left\{ \frac{i}{4} \int_0^s ds_1
\int_0^s ds_2  w(s_{1}) \cdot h^{-1}(s_{1}, s_{2}) \cdot w(s_{2})
\right\}} } \\ \nonumber & & \quad \times \left\{ m - i \gamma \cdot \left[
p + g^{2} \int{\frac{d^{4}k}{(2\pi)^{4}}  \, \int_{0}^{s}{ds' \, \mathbf{\tilde{D}}_{th}(k) \cdot
\left[ w'(s') - 2p \right]  e^{-i k \cdot [w(s) - w(s')] + 2ik \cdot
p (s-s')} } } \right] \right\} \\ \nonumber  & & \quad \times \exp{\left\{ \frac{i}{2}\, g^{2} \, \int{\frac{d^{4}k}{(2\pi)^{4}} \, \int_{0}^{s}{ds_{1} \int_{0}^{s}{ds_{2} \, e^{i k \cdot [w(s_{2})-w(s_{1})] - 2ik \cdot p (s_{2} - s_{1}) } \, \left[
w'(s_{1}) - 2 p \right] \cdot \mathbf{\tilde{D}}_{th}(k) \cdot \left[ w'(s_{2}) -
2 p \right] }}} \right\}},
\end{eqnarray}
\end{widetext}

\noindent where
\begin{eqnarray}\label{EqA-9}
& & h(s_{1},s_{2}) = \int_{0}^{s}{ds' \, \Theta(s_{1} - s') \Theta(s_{2} - s')}, \\ \nonumber & & h^{-1}(s_{1},s_{2}) = \frac{\partial}{\partial s_{2}} \frac{\partial}{\partial s_{1}} \delta(s_{1} - s_{2})
\end{eqnarray}

\noindent and
\begin{eqnarray}\label{EqA-10}
& & e^{+ \frac{1}{2} \mathrm{Tr}{\ln{\left( 2h \right)}}} = \int{d[w] \,} \\ \nonumber & &  \quad \cdot { \exp{\left\{ \frac{i}{4} \int_{0}^{s}{ds_{1} \, \int_{0}^{s}{ds_{2} \, w(s_{1}) \cdot h^{-1}(s_{1}, s_{2})\cdot
w(s_{2}) }} \right\}} }
\end{eqnarray}

\noindent The Bloch-Nordseick (BN) set of approximations is completed by the replacements, which relfect the neglect of momentum fluctuation of magnitude less than $p$, $[w(s') - 2s'p] \rightarrow -2s'p$ and $[w'(s') - 2p] \rightarrow -2p$, so as to get eventually the expression
\begin{eqnarray}\label{EqA-12}
& & {\langle \vec{p}, n| {\mathbf{S}'}^{BN}_{th} |\vec{y}, y_{0} \rangle}_{Q}
\\ \nonumber &\simeq&  (\mathbf{Z}_{0}[i \tau])^{-1} \left[(2\pi)^{3} \tau
\right]^{-1/2} e^{-i(\vec{p}\cdot\vec{y} - \omega_{n}y_{0})} \,
\left\{ m - i \gamma \cdot p  \right\} \\ \nonumber  & & \times \, i \int_{0}^{\infty}{ds \, e^{-is (m^2 + p^2)} } \\ \nonumber & & \quad \quad \cdot { e^{ {2i} g^{2} \int{\frac{d^{4}k}{(2\pi)^{4}} \, \int_{0}^{s}{ds_{1} \,
\int_{0}^{s}{ds_{2} \, [p \cdot {\mathbf{\tilde{D}}}_{th}(k) \cdot p] \, e^{2ik
\cdot p (s_{1} - s_{2})} }}} } }
\end{eqnarray}
which, in the main text, is at the level of Eqs. (\ref{Eq2-10})-(\ref{Eq2-11}). Note that in passing from Eq. (\ref{EqA-8}) to (\ref{EqA-12}), the huge parenthesis of Eq. (\ref{EqA-8}),
\begin{widetext}
\begin{equation}
\left\{ m - i \gamma \cdot \left[
p + g^{2} \int{\frac{d^{4}k}{(2\pi)^{4}} \, \int_{0}^{s}{ds' \, \mathbf{\tilde{D}}_{th}(k) \cdot
\left[ w'(s') - 2p \right]  e^{-i k \cdot [w(s) - w(s')] + 2ik \cdot
p (s-s')} } } \right] \right\} \nonumber
\end{equation}
\end{widetext}

\noindent has simply been replaced by $\{m - i \gamma \cdot p\}$. That is, the medium generated mass-shift, a long noticed phenomenon \cite{Bechler1981a, Cox1984a, Donoghue1985a}, has been withdrawn from our computation, because in this article, focus is placed on the depletion mechanisms acting on the incident fermion, and the subsequently-generated shock waves inside the thermalized medium.

\section{\label{appB}On the Doppler model}

We here propose an elementary, one-dimensional derivation which avoids the problematic
Lorentz transformation of the temperature $T$ \cite{tHW1971}, of the $p(z_0)$
fall-off. Let $\rho(\nu)$ be the linear density of photons per unit
length, at thermal equilibrium.  Then $\rho(\nu) \, \delta\nu$ is the
number of such photons, of energy $h \nu$, per unit of time, in the
frequency interval $\delta \nu$. The momentum carried by any photon
in that distribution is $h \nu /c$.  What the charged traveling
particle sees is a Doppler shift of frequencies: for the photons
"approaching head on", $\nu \rightarrow \nu_{+}$, and for the
photons "approaching from the rear", $\nu \rightarrow \nu_{-}$, with
\begin{equation}\label{EqB-1}
\nu_{+} = \nu \, \sqrt{\frac{c+v}{c-v}}, \quad \nu_{-} = \nu \,
\sqrt{\frac{c-v}{c+v}}
\end{equation}

\noindent From the elementary diagram of QED, let $\eta \alpha$ be
the absorption probability of a photon by the fermionic line, where
$\eta$ stands for some numerical constant, $\alpha$ for the fine
structure constant.

The number of photons absorbed per unit of time in the frequency
interval $\delta\nu$ is thus $\alpha \eta \rho(\nu) \, \delta\nu$, the
same in either front and rear directions.  This allows the
calculation of the momentum change induced by the process,
assuming that all the other interactions, with the heat bath
thermalized photons, average out to zero. One gets
\begin{equation}\label{EqB-2}
\frac{\textrm{d}p}{\textrm{d}t} = - \eta \alpha \int_{0}^{\infty}
{\textrm{d}\nu \, \rho(\nu) \, \frac{h \nu}{c} \,
\left({\sqrt{\frac{c+v}{c-v}}}-{\sqrt{\frac{c-v}{c+v}}}\right)},
\end{equation}

\noindent that is,
\begin{eqnarray}\label{EqB-3}
& & \frac{\textrm{d} p}{\textrm{d}t}=-p \ \Gamma_{\mathrm{Doppler}}, \\ \nonumber & & \quad
\Gamma_\mathrm{{Doppler}}=\left(\frac{\eta \pi^{2}}{3}\right) \, \frac{\alpha c}{\lambda_{c}} \, \left(\frac{k_{B} T}{m c^{2}}\right)^{2},
\end{eqnarray}

\noindent and this is the $\Gamma$-constant that appears in Section
\ref{sec2}, Eq.~(\ref{Eq2-23}).

\section{\label{appC}Approximation Scheme in the $s$-integral}

The integral in the exponential factor of Eq.~(\ref{Eq2-11}) becomes,
\begin{widetext}
\begin{equation}\label{EqC-1}
- \frac{g^{2}}{2 \pi^{3}} \, \vec{p}^{\, 2} \, \int_{0}^{s}{ds_{1} \,
\int_{0}^{s_{1}}{ds' \, \int{d\Omega \, (1 - \zeta^{2}) \,
\int_{0}^{\infty}{dk \, k \, \frac{\cos( 2 k \omega_{n} s') \, \cos( 2
k p \zeta s')}{e^{\beta k} - 1} \, e^{-\frac{k}{p}} } }}},
\end{equation}

\noindent where $\zeta = \cos{\theta}$.  First carrying out both the
$s_{1}$- and $s'$-integrals, it reduces to
\begin{equation}\label{EqC-2}
- \frac{ g^{2}}{8 \pi^{3}} \, \vec{p}^{\, 2} \, \int{d\Omega \,
(1-\zeta^{2}) \, \int_{0}^{\infty}{dk \, \frac{k \, e^{-k/p}}{(e^{\beta
k} - 1)} \, \left[ \frac{1 - \cos(2 s Q^{(+)})}{(Q^{(+)})^{2}} +
\frac{1 - \cos(2 s Q^{(-)})}{(Q^{(-)})^{2}}\right] } },
\end{equation}
\end{widetext}

\noindent where $Q^{(\pm)} = k ( p \zeta \pm \omega_{n})$. The
integral over $k$ is a bit complicated and can not be carried out
exactly.  To continue evaluation, observe the oscillating factor
$\exp{[-is(\omega^{2} - \omega_{n}^{2}) ]}$ in the $s$-integral of
Eq. (\ref{Eq2-10}); when $s > s_{max} = (\omega^{2} -
\omega_{n}^{2})^{-1}$, the oscillating factor effectively removes any contribution.  The arguments of the cosine factors are
\begin{equation}\label{EqC-3}
\nonumber |s Q^{(\pm)}| < s_{max} k |p \zeta \pm \omega_{n}| \ll s_{max} p |p
\zeta \pm \omega_{n}| < \frac{p |p \zeta \pm \omega_{n}|}{\omega^{2}
- \omega_{n}^{2}}.
\end{equation}

\noindent Since $p \gg m$ and $|\zeta| < 1$,
\begin{equation}\label{EqC-4}
\nonumber |sQ^{(\pm)}| \ll \frac{p}{|p \mp \omega_{n}|} = \frac{1}{|1 \mp
\frac{\omega_{n}}{p}|} = \frac{1}{|1 \pm i (2n+1) \pi
\frac{T}{p}|},
\end{equation}

\noindent or effectively $|s Q^{(\pm)}| < 1$.  Thus, the arguments of the cosine functions are small and can be approximated as
\begin{equation}\label{EqC-5}
\frac{1 - \cos(2 s Q^{(\pm)})}{(Q^{(\pm)})^{2}} \simeq \frac{1}{2}
\frac{(2 s Q^{(\pm)})^{2}}{(Q^{(\pm)})^{2}} =  2 s^{2}.
\end{equation}

\noindent Set $x=k/p$ and the $k$-integral becomes
\begin{equation}\label{EqC-6}
\nonumber \int_{0}^{\infty}{dk \, \frac{k \, e^{-k/p}}{e^{\beta k} - 1}} =
\vec{p}^{\, 2} \, \int_{0}^{\infty}{dx \, \frac{x \, e^{-x}}{e^{x(p/T)} -
1}} \equiv \vec{p}^{\, 2} \, f(\frac{T}{p});
\end{equation}

\noindent which yields
\begin{eqnarray}\label{EqC-7}
\nonumber & & - s^{2} \frac{g^{2}}{2 \pi^{3}} \, (\vec{p}^{\, 2})^{2} \,
\int_{0}^{1}{d\zeta \, (1-\zeta^{2}) \, \int_{0}^{\infty}{dx \,
\frac{x \, e^{-x}}{e^{x (T/p)} - 1}}} \\ &=& - \xi^{2} s^{2} g^{2}
f(\frac{T}{p}) (\vec{p}^{\, 2})^{2},
\end{eqnarray}

\noindent where all numerical factors have been combined into $\xi^{2} =
\frac{4}{3\pi}$.  Similar approximations are made to derive Eqs. (\ref{Eq3-10}) and
(\ref{Eq3-16}).

%
%
\bibliography{QED-FT_03152008}

\begin{thebibliography}{22}
\expandafter\ifx\csname natexlab\endcsname\relax\def\natexlab#1{#1}\fi
\expandafter\ifx\csname bibnamefont\endcsname\relax
  \def\bibnamefont#1{#1}\fi
\expandafter\ifx\csname bibfnamefont\endcsname\relax
  \def\bibfnamefont#1{#1}\fi
\expandafter\ifx\csname citenamefont\endcsname\relax
  \def\citenamefont#1{#1}\fi
\expandafter\ifx\csname url\endcsname\relax
  \def\url#1{\texttt{#1}}\fi
\expandafter\ifx\csname urlprefix\endcsname\relax\def\urlprefix{URL }\fi
\providecommand{\bibinfo}[2]{#2}
\providecommand{\eprint}[2][]{\url{#2}}

\bibitem[{\citenamefont{Candelpergher et~al.}(2005)\citenamefont{Candelpergher,
  Fried, and Grandou}}]{CFG2005}
\bibinfo{author}{\bibfnamefont{B.}~\bibnamefont{Candelpergher}},
  \bibinfo{author}{\bibfnamefont{H.~M.} \bibnamefont{Fried}}, \bibnamefont{and}
  \bibinfo{author}{\bibfnamefont{T.}~\bibnamefont{Grandou}},
  \bibinfo{journal}{Int.\ J.\ Mod.\ Phys.} \textbf{\bibinfo{volume}{20}},
  \bibinfo{pages}{7525} (\bibinfo{year}{2005}).

\bibitem[{\citenamefont{Fried}(1972)}]{HMF1972}
\bibinfo{author}{\bibfnamefont{H.~M.} \bibnamefont{Fried}},
  \emph{\bibinfo{title}{Functional Methods and Models in Quantum Field Theory}}
  (\bibinfo{publisher}{The MIT Press}, \bibinfo{address}{Cambridge, MA},
  \bibinfo{year}{1972}).

\bibitem[{\citenamefont{Fried}(1990)}]{HMF1990}
\bibinfo{author}{\bibfnamefont{H.~M.} \bibnamefont{Fried}},
  \emph{\bibinfo{title}{Basics of Functional Methods and Eikonal Models}}
  (\bibinfo{publisher}{Editions Fronti\`{e}res},
  \bibinfo{address}{Gif-sur-Yvette Cedex, France}, \bibinfo{year}{1990}).

\bibitem[{\citenamefont{Fried}(2002)}]{HMF2002}
\bibinfo{author}{\bibfnamefont{H.~M.} \bibnamefont{Fried}},
  \emph{\bibinfo{title}{Green's Functions and Ordered Exponentials}}
  (\bibinfo{publisher}{Cambridge University Press},
  \bibinfo{address}{Cambridge}, \bibinfo{year}{2002}).

\bibitem[{\citenamefont{Weldon}(1991)}]{Weldon1991a}
\bibinfo{author}{\bibfnamefont{H.~A.} \bibnamefont{Weldon}},
  \bibinfo{journal}{Phys.\ Rev. D} \textbf{\bibinfo{volume}{44}},
  \bibinfo{pages}{3955} (\bibinfo{year}{1991}).

\bibitem[{\citenamefont{Takashiba}(1996)}]{Takashiba1996a}
\bibinfo{author}{\bibfnamefont{K.}~\bibnamefont{Takashiba}},
  \bibinfo{journal}{Int. J. Mod. Phys. A} \textbf{\bibinfo{volume}{11}},
  \bibinfo{pages}{2309} (\bibinfo{year}{1996}).

\bibitem[{\citenamefont{Blaizot and Iancu}(1996)}]{BI1996a}
\bibinfo{author}{\bibfnamefont{J.~P.} \bibnamefont{Blaizot}} \bibnamefont{and}
  \bibinfo{author}{\bibfnamefont{E.}~\bibnamefont{Iancu}},
  \bibinfo{journal}{Phys.\ Rev. Lett.} \textbf{\bibinfo{volume}{76}},
  \bibinfo{pages}{3080} (\bibinfo{year}{1996}).

\bibitem[{\citenamefont{Blaizot and Iancu}(1997{\natexlab{a}})}]{BI1997a}
\bibinfo{author}{\bibfnamefont{J.~P.} \bibnamefont{Blaizot}} \bibnamefont{and}
  \bibinfo{author}{\bibfnamefont{E.}~\bibnamefont{Iancu}},
  \bibinfo{journal}{Phys.\ Rev. D} \textbf{\bibinfo{volume}{55}},
  \bibinfo{pages}{973} (\bibinfo{year}{1997}{\natexlab{a}}).

\bibitem[{\citenamefont{Blaizot and Iancu}(1997{\natexlab{b}})}]{BI1997b}
\bibinfo{author}{\bibfnamefont{J.~P.} \bibnamefont{Blaizot}} \bibnamefont{and}
  \bibinfo{author}{\bibfnamefont{E.}~\bibnamefont{Iancu}},
  \bibinfo{journal}{Phys.\ Rev. D} \textbf{\bibinfo{volume}{56}},
  \bibinfo{pages}{7877} (\bibinfo{year}{1997}{\natexlab{b}}).

\bibitem[{\citenamefont{Sheu}(2008)}]{Sheu2008a}
\bibinfo{author}{\bibfnamefont{Y.-M.} \bibnamefont{Sheu}}, Ph.D. thesis,
  \bibinfo{school}{Brown University}, \bibinfo{address}{Providence, RI, USA}
  (\bibinfo{year}{2008}).

\bibitem[{\citenamefont{{Le Bellac}}(2000)}]{LeBellac2000}
\bibinfo{author}{\bibfnamefont{M.}~\bibnamefont{{Le Bellac}}},
  \emph{\bibinfo{title}{Thermal Field Theory}} (\bibinfo{publisher}{Cambridge
  University Press}, \bibinfo{address}{Cambridge, UK}, \bibinfo{year}{2000}).

\bibitem[{\citenamefont{Corless et~al.}(1996)\citenamefont{Corless, Gonnet,
  Hare, Jeffrey, and Knuth}}]{Lambert}
\bibinfo{author}{\bibfnamefont{R.~M.} \bibnamefont{Corless}},
  \bibinfo{author}{\bibfnamefont{G.~H.} \bibnamefont{Gonnet}},
  \bibinfo{author}{\bibfnamefont{D.~E.~G.} \bibnamefont{Hare}},
  \bibinfo{author}{\bibfnamefont{D.~J.} \bibnamefont{Jeffrey}},
  \bibnamefont{and} \bibinfo{author}{\bibfnamefont{D.~E.} \bibnamefont{Knuth}},
  \bibinfo{journal}{Adv. Computational Math.} \textbf{\bibinfo{volume}{5}},
  \bibinfo{pages}{329} (\bibinfo{year}{1996}).

\bibitem[{\citenamefont{Grandou and Reynaud}(1997)}]{TGPR1997}
\bibinfo{author}{\bibfnamefont{T.}~\bibnamefont{Grandou}} \bibnamefont{and}
  \bibinfo{author}{\bibfnamefont{P.}~\bibnamefont{Reynaud}},
  \bibinfo{journal}{Nucl.\ Phys.\ B} \textbf{\bibinfo{volume}{486}},
  \bibinfo{pages}{164} (\bibinfo{year}{1997}).

\bibitem[{\citenamefont{Candelpergher and Grandou}(2000)}]{CG2005a}
\bibinfo{author}{\bibfnamefont{B.}~\bibnamefont{Candelpergher}}
  \bibnamefont{and} \bibinfo{author}{\bibfnamefont{T.}~\bibnamefont{Grandou}},
  \bibinfo{journal}{Ann. Phys.} \textbf{\bibinfo{volume}{283}},
  \bibinfo{pages}{232} (\bibinfo{year}{2000}).

\bibitem[{\citenamefont{Braaten and
  Pisarski}(1990{\natexlab{a}})}]{Braaten1990a}
\bibinfo{author}{\bibfnamefont{E.}~\bibnamefont{Braaten}} \bibnamefont{and}
  \bibinfo{author}{\bibfnamefont{R.~D.} \bibnamefont{Pisarski}},
  \bibinfo{journal}{Phys. Rev. Lett.} \textbf{\bibinfo{volume}{64}},
  \bibinfo{pages}{1338} (\bibinfo{year}{1990}{\natexlab{a}}).

\bibitem[{\citenamefont{Braaten and
  Pisarski}(1990{\natexlab{b}})}]{Braaten1990b}
\bibinfo{author}{\bibfnamefont{E.}~\bibnamefont{Braaten}} \bibnamefont{and}
  \bibinfo{author}{\bibfnamefont{R.~D.} \bibnamefont{Pisarski}},
  \bibinfo{journal}{Nucl.\ Phys.\ B} \textbf{\bibinfo{volume}{337}},
  \bibinfo{pages}{569} (\bibinfo{year}{1990}{\natexlab{b}}).

\bibitem[{\citenamefont{Braaten and
  Pisarski}(1990{\natexlab{c}})}]{Braaten1990c}
\bibinfo{author}{\bibfnamefont{E.}~\bibnamefont{Braaten}} \bibnamefont{and}
  \bibinfo{author}{\bibfnamefont{R.~D.} \bibnamefont{Pisarski}},
  \bibinfo{journal}{Nucl.\ Phys.\ B} \textbf{\bibinfo{volume}{339}},
  \bibinfo{pages}{310} (\bibinfo{year}{1990}{\natexlab{c}}).

\bibitem[{\citenamefont{Frenkel and Taylor}(1990)}]{Frenkel1990a}
\bibinfo{author}{\bibfnamefont{J.}~\bibnamefont{Frenkel}} \bibnamefont{and}
  \bibinfo{author}{\bibfnamefont{J.~C.} \bibnamefont{Taylor}},
  \bibinfo{journal}{Nucl.\ Phys.\ B} \textbf{\bibinfo{volume}{334}},
  \bibinfo{pages}{199} (\bibinfo{year}{1990}).

\bibitem[{\citenamefont{Bechler}(1981)}]{Bechler1981a}
\bibinfo{author}{\bibfnamefont{A.}~\bibnamefont{Bechler}},
  \bibinfo{journal}{Ann. Phys.} \textbf{\bibinfo{volume}{135}},
  \bibinfo{pages}{19} (\bibinfo{year}{1981}).

\bibitem[{\citenamefont{Cox et~al.}(1984)\citenamefont{Cox, Hellman, and
  Yildiz}}]{Cox1984a}
\bibinfo{author}{\bibfnamefont{P.~H.} \bibnamefont{Cox}},
  \bibinfo{author}{\bibfnamefont{W.~S.} \bibnamefont{Hellman}},
  \bibnamefont{and} \bibinfo{author}{\bibfnamefont{A.}~\bibnamefont{Yildiz}},
  \bibinfo{journal}{Ann. Phys.} \textbf{\bibinfo{volume}{154}},
  \bibinfo{pages}{211} (\bibinfo{year}{1984}).

\bibitem[{\citenamefont{Donoghue et~al.}(1985)\citenamefont{Donoghue, Holstein,
  and Robinett}}]{Donoghue1985a}
\bibinfo{author}{\bibfnamefont{J.~F.} \bibnamefont{Donoghue}},
  \bibinfo{author}{\bibfnamefont{B.~R.} \bibnamefont{Holstein}},
  \bibnamefont{and} \bibinfo{author}{\bibfnamefont{R.~W.}
  \bibnamefont{Robinett}}, \bibinfo{journal}{Ann. Phys. (NY)}
  \textbf{\bibinfo{volume}{164}}, \bibinfo{pages}{233} (\bibinfo{year}{1985}).

\bibitem[{\citenamefont{ter Haar and Wergeland}(1971)}]{tHW1971}
\bibinfo{author}{\bibfnamefont{D.}~\bibnamefont{ter Haar}} \bibnamefont{and}
  \bibinfo{author}{\bibfnamefont{H.}~\bibnamefont{Wergeland}},
  \bibinfo{journal}{Phys.\ Rep.} \textbf{\bibinfo{volume}{1C}},
  \bibinfo{pages}{31} (\bibinfo{year}{1971}).

\end{thebibliography}

\end{document}